\documentclass[]{spie}  
\usepackage[font=small]{caption}
\usepackage{amsmath,amsfonts,amssymb,gensymb}
\usepackage{graphicx}
\usepackage{subfig}
\usepackage{xspace} 

\def\spider{{\sc Spider}\xspace}

\def\planck{{\it Planck}\xspace}
\def\bicep{{\sc Bicep}\xspace}

\title{Design and pre-flight performance of SPIDER 280~GHz receivers}

\author[a]{E.C. ~Shaw}
\author[b]{P.A.R. ~Ade}
\author[c]{S. ~Akers}
\author[d]{M. ~Amiri}
\author[e]{J. ~Austermann}
\author[e]{J. ~Beall}
\author[e]{D.T. ~Becker}
\author[f]{S.J. ~Benton}
\author[f]{A.S. ~Bergman}
\author[g,h]{J.J. ~Bock}
\author[i]{J.R. ~Bond}
\author[j]{S.A. ~Bryan}
\author[k]{H.C. ~Chiang}
\author[l]{C.R. ~Contaldi}
\author[m]{R.S. ~Domagalski}
\author[g,h]{O. ~Dor\'e}
\author[e]{S.M. ~Duff}
\author[f]{A.J. ~Duivenvoorden}
\author[n]{H.K. ~Eriksen}
\author[o]{M. ~Farhang}
\author[a,p]{J.P. ~Filippini}
\author[q]{L.M. ~Fissel}
\author[f]{A.A. ~Fraisse}
\author[r,s]{K. ~Freese}
\author[n]{M. ~Galloway}
\author[t]{A.E. ~Gambrel}
\author[m]{N.N. Gandilo}
\author[u]{K. ~Ganga}
\author[e]{A. ~Grigorian}
\author[v]{R. ~Gualtieri}
\author[s]{J.E. ~Gudmundsson}
\author[d]{M. ~Halpern}
\author[m]{J. ~Hartley}
\author[w]{M. ~Hasselfield}
\author[e]{G. ~Hilton}
\author[h]{W. ~Holmes}
\author[g]{V.V. ~Hristov}
\author[i]{Z. ~Huang}
\author[e]{J. ~Hubmayr}
\author[x,y]{K.D. ~Irwin}
\author[f]{W.C. ~Jones}
\author[a]{A. ~Kahn}
\author[x]{C.L. ~Kuo}
\author[f]{Z.D. ~Kermish}
\author[a]{A. ~Lennox}
\author[m,z]{J.S.-Y. ~Leung}
\author[f]{S. ~Li}
\author[g]{P.V. ~Mason}
\author[h]{K. ~Megerian}
\author[n]{L.M. Mocanu}
\author[g]{L. ~Moncelsi}
\author[g]{T.A. ~Morford}
\author[1,2]{J.M. ~Nagy}
\author[a]{R. ~Nie}
\author[m,3,z]{C.B. ~Netterfield}
\author[i]{M. ~Nolta}
\author[a]{B. ~Osherson}
\author[3,4]{I.L. ~Padilla}
\author[t,5]{A.S. ~Rahlin}
\author[6]{S. ~Redmond}
\author[e]{C. ~Reintsema}
\author[7]{L.J. ~Romualdez}
\author[c]{J.E. ~Ruhl}
\author[h]{M.C. ~Runyan}
\author[i]{J.A. ~Shariff}
\author[f]{C. ~Shiu}
\author[8,9]{J.D. ~Soler}
\author[f]{X. ~Song}
\author[n]{H. Thommesen}
\author[g,h]{A. ~Trangsrud}
\author[b]{C. ~Tucker}
\author[g]{R.S. ~Tucker}
\author[h]{A.D. ~Turner}
\author[e]{J. ~Ullom}
\author[f]{J.F. ~van der List}
\author[e]{J. ~Van Lanen}
\author[e]{M.R. ~Vissers}
\author[h]{A.C. ~Weber}
\author[c]{S. ~Wen}
\author[n]{I.K. ~Wehus}
\author[d]{D.V. ~Wiebe}
\author[x,10]{E.Y. ~Young}
\affil[a]{Department of Physics, University of Illinois at Urbana-Champaign, Urbana, IL, USA}
\affil[b]{School of Physics and Astronomy, Cardiff University, Cardiff, UK}
\affil[c]{Physics Department, Case Western Reserve University, Cleveland, OH, USA}
\affil[d]{Department of Physics and Astronomy, University of British Columbia, Vancouver, BC, Canada}
\affil[e]{National Institute of Standards and Technology, Boulder, CO, USA}
\affil[f]{Department of Physics, Princeton University, Princeton, NJ, USA}
\affil[g]{Division of Physics, Mathematics and Astronomy, California Institute of Technology, Pasadena, CA, USA}
\affil[h]{Jet Propulsion Laboratory, Pasadena, CA, USA}
\affil[i]{Canadian Institute for Theoretical Astrophysics, University of Toronto, Toronto, ON, Canada}
\affil[j]{School of Electrical, Computer, and Energy Engineering, Arizona State University, Tempe, AZ, USA}
\affil[k]{Department of Physics, McGill University, Montreal, QC, Canada}
\affil[l]{Blackett Laboratory, Imperial College London, London, UK}
\affil[m]{Department of Astronomy and Astrophysics, University of Toronto, Toronto, ON, Canada}
\affil[n]{Institute of Theoretical Astrophysics, University of Oslo, Oslo, Norway}
\affil[o]{Department of Physics, Shahid Beheshti University, Velenjak, Tehran, Iran}
\affil[p]{Department of Astronomy, University of Illinois at Urbana-Champaign, Urbana, IL, USA}
\affil[q]{Department of Physics, Engineering Physics and Astronomy, Queen's University, Kingston, ON, Canada}
\affil[r]{Department of Physics, The University of Texas at Austin, Austin, TX, USA}
\affil[s]{The Oskar Klein Centre for Cosmoparticle Physics, Department of Physics, Stockholm University, SE-106 91 Stockholm, Sweden}
\affil[t]{Kavli Institute for Cosmological Physics, University of Chicago, Chicago, IL, USA}
\affil[u]{Universit\'e de Paris, CNRS, AstroParticule et Cosmologie, F-75013 Paris, France}
\affil[v]{HEP, Argonne National Laboratory, Lemont, IL, USA}
\affil[w]{Department of Astronomy and Astrophysics, Pennsylvania State University, University Park, PA, USA}
\affil[x]{Department of Physics, Stanford University, Stanford, CA, USA}
\affil[y]{SLAC National Accelerator Laboratory, Menlo Park, CA, USA}
\affil[z]{Dunlap Institute for Astronomy and Astrophysics, University of Toronto, Toronto, Ontario M5S 3H4, Canada}
\affil[1]{Department of Physics, Washington University in St. Louis, St. Louis, MO, USA}
\affil[2]{McDonnell Center for the Space Sciences, Washington University in St. Louis, St. Louis, MO, USA}
\affil[3]{Department of Physics, University of Toronto, Toronto, ON, Canada}
\affil[4]{Department of Physics and Astronomy, Johns Hopkins University, Baltimore, MD, USA}
\affil[5]{Fermi National Accelerator Laboratory, Batavia, IL, USA}
\affil[6]{Department of Mechanical and Aerospace Engineering, Princeton University, Princeton, NJ, USA}
\affil[7]{University of Toronto Institute for Aerospace Studies, Toronto, ON, Canada}
\affil[8]{Max-Planck-Institute for Astronomy, Heidelberg, Germany}
\affil[9]{Laboratoire AIM, Paris-Saclay, CEA/IRFU/SAp - CNRS - Universit\'e Paris Diderot, 91191, Gif-sur-Yvette Cedex, France}
\affil[10]{Kavli Institute for Particle Astrophysics and Cosmology, Menlo Park, CA, USA}

\authorinfo{E-mail: ecshaw2@illinois.edu, Telephone: 1 217 300 0300}

\begin{document} 
\maketitle

\begin{abstract}
In this work we describe upgrades to the \spider balloon-borne telescope in preparation for its second flight, currently planned for December 2021. The \spider instrument is optimized to search for a primordial B-mode polarization signature in the cosmic microwave background at degree angular scales.
During its first flight in 2015, \spider mapped $\sim 10 \%$ of the sky at 95 and 150~GHz. 
The payload for the second Antarctic flight will incorporate three new 280~GHz receivers alongside three refurbished 95- and 150~GHz receivers from \spider's first flight. 
In this work we discuss the design and characterization of these new receivers, which employ over 1500 feedhorn-coupled transition-edge sensors. 
We describe pre-flight laboratory measurements of detector properties, and the optical performance of completed receivers.
These receivers will map a wide area of the sky at 280 GHz, providing new information on polarized Galactic dust emission that will help to separate it from the cosmological signal.
\end{abstract}


\section{INTRODUCTION}
\label{sec:intro}  
\spider is a balloon-borne millimeter-wave polarimeter specifically designed to scan the cosmic microwave background (CMB) for B-mode polarization patterns at degree angular scales, a unique signature of primordial gravitational waves (PGWs), and to separate this faint CMB signal from polarized Galactic emission \cite{Filippini_2010,  Ade_2015, Fraisse_2013,Gualtieri_2018}.
If detected, this signal would provide insight into early-universe physics at energies far beyond those accessible to particle accelerators~\cite{abazajian2016cmbs4}.
The \spider experiment seeks to either detect or set an upper limit on the tensor-to-scalar ratio, $r$, below 0.03 at a $3\sigma$ confidence level \cite{Fraisse_2013, Rahlin_2014}.

The primordial B-mode polarization signal is predicted to be faint, with anisotropies $\ll 1~\mu \text{K}$.
Experimental detection requires extremely sensitive detectors and tight control over polarized instrumental systematics. 
Detection is further complicated by the presence of foreground polarized and unpolarized emission from Galactic and atmospheric sources.
Polarized foreground emission is minimal at $\sim$70~GHz across the full sky~\cite{collaboration2018planck}, and is dominated by synchrotron emission at lower frequencies and by thermal emission from small dust grains aligned with the Galactic magnetic field at higher frequencies.  
On angular scales where the primordial B-mode signal is expected to peak, the polarized CMB signal is obscured by polarized foregrounds at all frequencies.
The ability to characterize and remove these polarized foregrounds is thus crucial to recovering PGW B-modes, and demands deep observations at multiple frequencies.

The high ($\gtrsim250$~GHz) frequencies that are most effective for characterizing Galactic dust emission spectrum are challenging from the ground due to atmospheric opacity.
Atmospheric fluctuations and turbulence muddle the sky signal at large angular scales from the ground, often requiring time stream filtering that suppresses these modes.
At higher altitudes these atmospheric effects are diminished, and experimental platforms can take advantage of current technology to push detector sensitivity toward the CMB photon noise limit.

\spider observes from the stratosphere at an altitude of $\sim$36~km, above 99$\%$ of the atmosphere, on a NASA Long Duration Balloon (LDB).
In January 2015 \spider completed its first flight around the South Pole from the LDB Facility at McMurdo Station, Antarctica. 
During its 16~day flight \spider mapped as large an area as possible while avoiding the Galaxy and Sun (roughly 10$\%$ of the sky) with three telescopes tuned to 95~GHz and three to 150~GHz. 
The second \spider campaign, named ``\spider-2", will observe the same sky region and will redeploy two 95~GHz telescopes and one 150~GHz telescope along with three new 280~GHz telescopes\cite{Gualtieri_2018}.
The 280~GHz frequency band, the frequency at which the sky emission is dominated by Galactic dust, will provide an independent and deeper map compared to the 217 and 353~GHz \planck-HFI maps already available\cite{collaboration2018planckoverview}.
The obtained dust map will be used to extrapolate the foreground contamination to lower frequencies.

In this proceeding we describe updates to the \spider-2 payload, as well as the design and characterization of the new 280~GHz telescopes.
In Section \ref{section:Payload} we discuss the \spider payload and cryogenic design.
The \spider receiver architecture and 280~GHz receiver design is discussed in Section \ref{section:Receivers}, and the instrument detector and optical performance is discussed in Section \ref{section:InstrumentPerformance}.

\section{Spider Payload and Cryogenic Design}\label{section:Payload}
\spider's six monochromatic, on-axis, refracting telescopes are housed in a shared-vacuum, 1300~L, liquid helium cryostat \cite{Gudmundsson_2015}. 
The main tank (MT) provides 4~K cooling power to the telescope receivers, which extend up through the tank's cylindrical ports, and to the cryogenic half-wave plates.
The MT is surrounded by two stages of vapor-cooled shields, VCS1 and VCS2, that reduce the load on the helium tanks. 
Helium vapor from the MT passes through tubing in the VCSs, cooling the shields and a series of metal-mesh radiation filters~\cite{metal_mesh_review} mounted to the VCS shells.
The gaps between each stage are filled with multi-layer insulation (MLI) to reduce radiative loading. 
The telescopes' internal optical baffling and sub-kelvin refrigerators are serviced by a 20~L superfluid tank that is filled continuously from the MT through capillaries.
The superfluid tank is pumped down before launch and maintained by a small diaphragm pump during ascent; once the payload reaches float altitude it is vented to the vacuum of near-space, providing a stable temperature of 1.6~K.
Closed-cycle helium-3 sorption refrigerators inside each receiver cool the focal planes and detectors assemblies to their 300~mK operating temperatures.

\begin{figure}[ht]
\centering
\begin{minipage}[ht]{.495\linewidth}
\begin{center}
    \includegraphics[width=\textwidth]{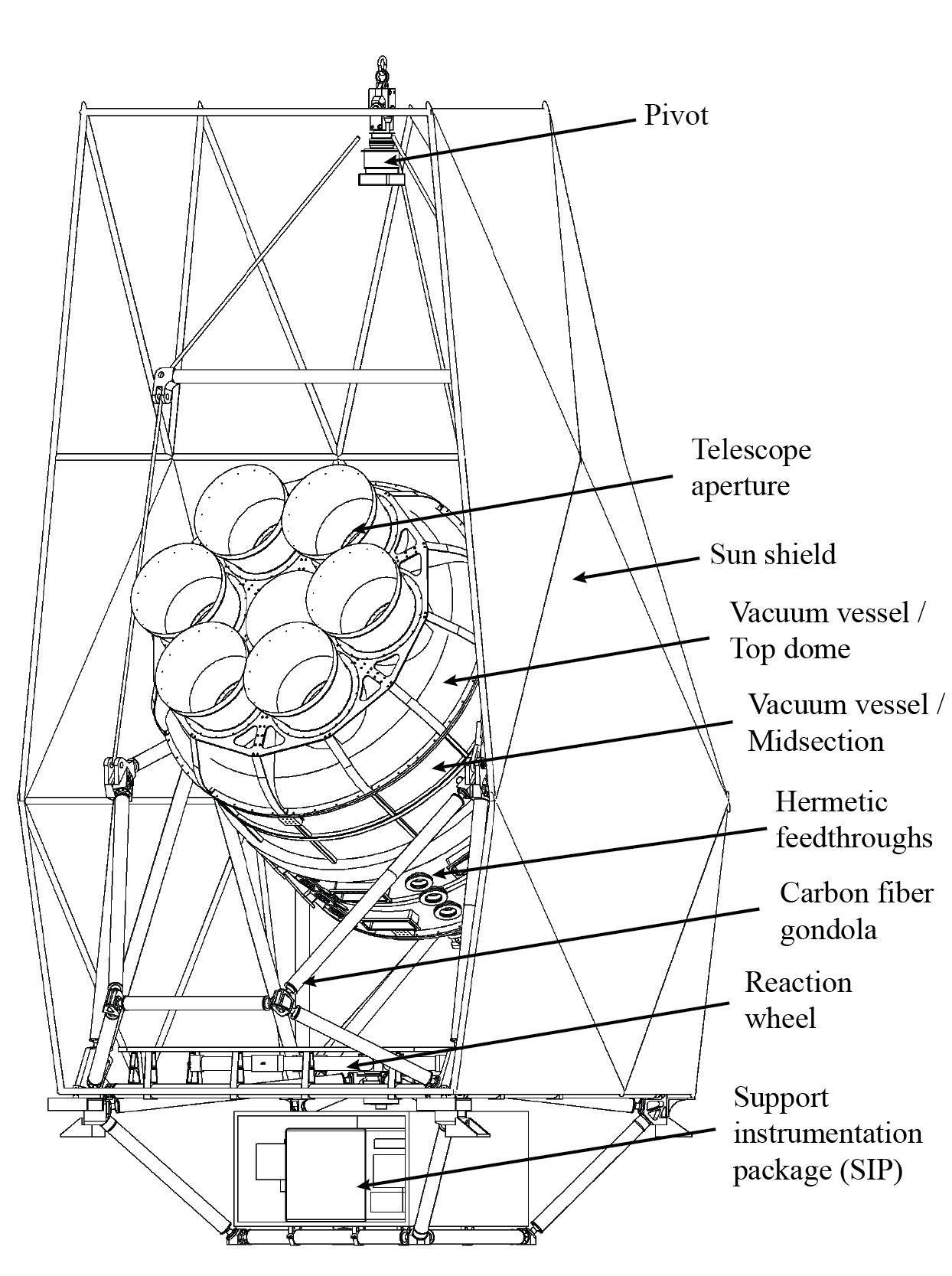}
\end{center}
\end{minipage}\hfill
\begin{minipage}[ht]{.45\linewidth}
\begin{center}
    \includegraphics[width=.95\textwidth]{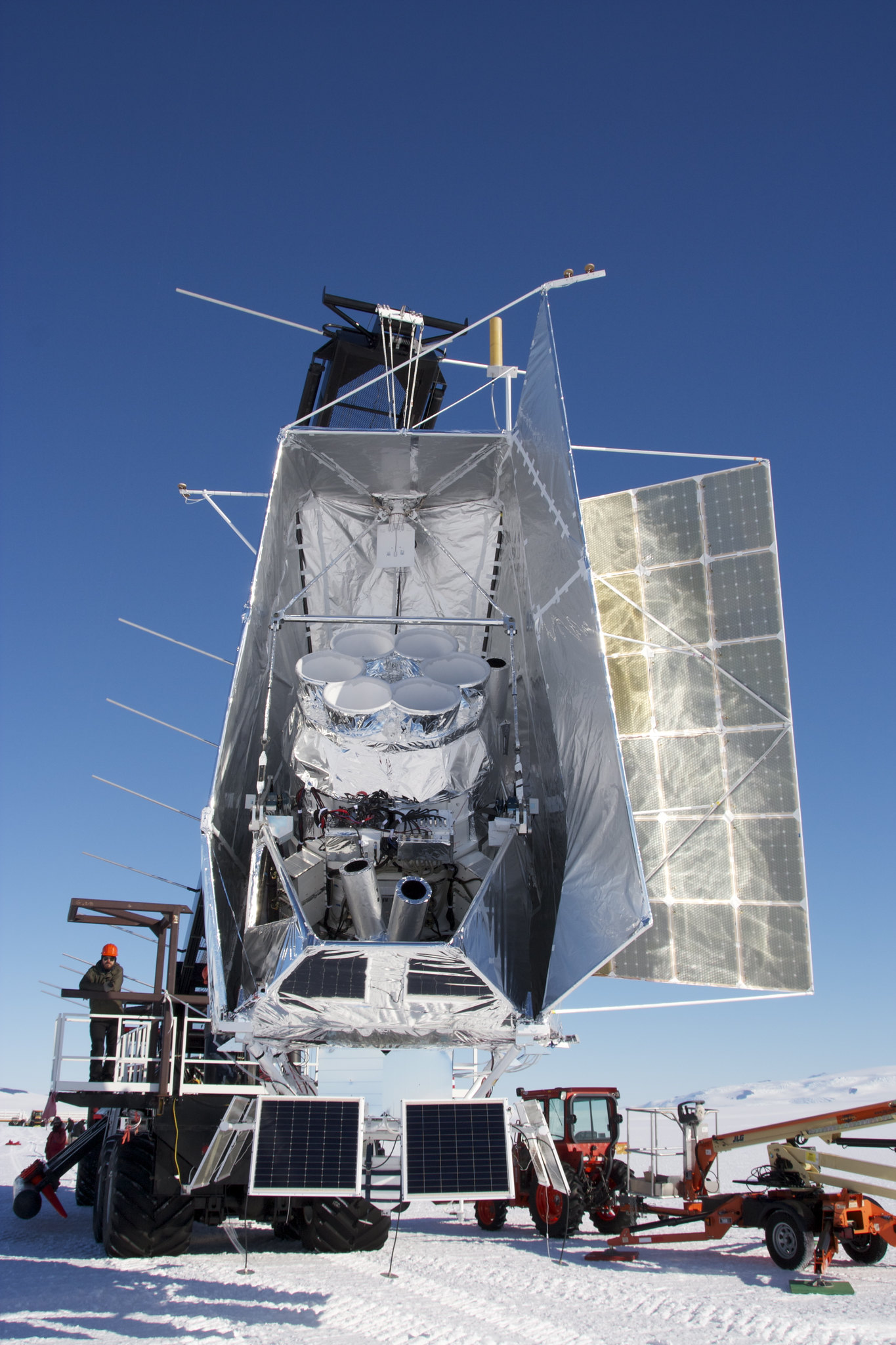}
\end{center}
\end{minipage}
\caption{
\textit{(Left)}: A simplified CAD model of the \spider payload, including the outer vacuum vessel, gondola, sun-shields, and Columbia Scientific Balloon Facility Support Instrumentation Package (SIP). 
The sun-shields are composed of a lightweight carbon fiber frame which is tiled with foam and aluminized Mylar. The SIP houses antenna power and control systems, and is mounted underneath the cryostat at the bottom of the payload. Reproduced from Ref.~\citenum{Gudmundsson_2015}.
\textit{(Right)}: Fully integrated \textsc{Spider} payload and balloon on the launchpad, prepared for launch in January 2015. Some key components visible in the photograph that are omitted in the CAD model are the solar panels, star cameras, and electronics.}
\label{payload_and_insert}
\end{figure}

The cryostat, readout electronics, pointing controls, and solar panel arrays are framed and supported by a lightweight carbon fiber and aluminum gondola \cite{Shariff_2014,Soler_2014}. 
The instrument is scanned in elevation using dual linear  actuators, and in azimuth using a motorized reaction wheel and pivot motor.
The payload is protected from the 24-hour antarctic summer daylight by Sun shields wrapped around the frame, and is shielded from glint off the Earth's surface by forebaffles extending from the telescope apertures.
Figure~\ref{payload_and_insert} shows the fully integrated \spider-1 payload and balloon on the launchpad at the McMurdo LDB facility.

All \spider hardware was recovered from the Antarctic ice in late 2015.
\spider-2 utilizes a similar design and will reuse much of the same hardware as \spider-1, but will fly with upgrades to the gondola and cryostat.
The gondola frame and sun-shield are entirely rebuilt for the second flight, with several changes to the overall design to reduce mass and improve strength.
The gondola will include upgraded gyroscopes and dual boresight star cameras that will provide the required in-flight pointing control and post-flight reconstruction with reduced mass and complexity relative to the first flight. 
For the second flight we have fabricated a new flight cryostat based on the previous design, manufactured at Meyer Tool\footnote{\textit{Meyer Tool \& Mfg}, Oak Lawn, IL.}.
The new flight cryostat includes a larger liquid helium tank, and has demonstrated improved vacuum performance relative to the original unit~\cite{Gudmundsson_2015}.
The first flight observed additional heat load on cryogenic stages and superfluid film build-up on the focal planes that were ascribed to a known helium leak; these have not been observed during multiple cool-downs of the new cryostat.
Taken together, the new system is expected to provide for a substantially longer hold time, and thus a longer flight.


\section{Spider Receivers}\label{section:Receivers}
Each \spider receiver is a small-aperture (270~mm optical stop diameter), two-lens, refracting telescope.
The design for the receivers is similar to that of \bicep~\cite{YoonBICEP}, with modifications for the weight and size constraints of a ballooning experiment, as well as for the lower loading environment at 36~km altitude. 
The on-axis design of the telescopes reduces polarized systematics due to instrument asymmetry. The highly modular architecture allows \spider to make observations at multiple frequencies with limited changes to the main infrastructure by simply swapping telescopes between receiver ports. 
\spider-2 will fly two 95~GHz and one 150~GHz receivers recovered from the first flight.
The telescopes chosen for \spider-2 are selected for their high performance during the first flight.
The receivers have been inspected in the laboratory post-flight, tested in the new flight cryostat, and are ready for another launch. 

\begin{figure}[ht]
\centering
\begin{minipage}[ht]{.6\linewidth}
\begin{center}
    \includegraphics[height=13cm]{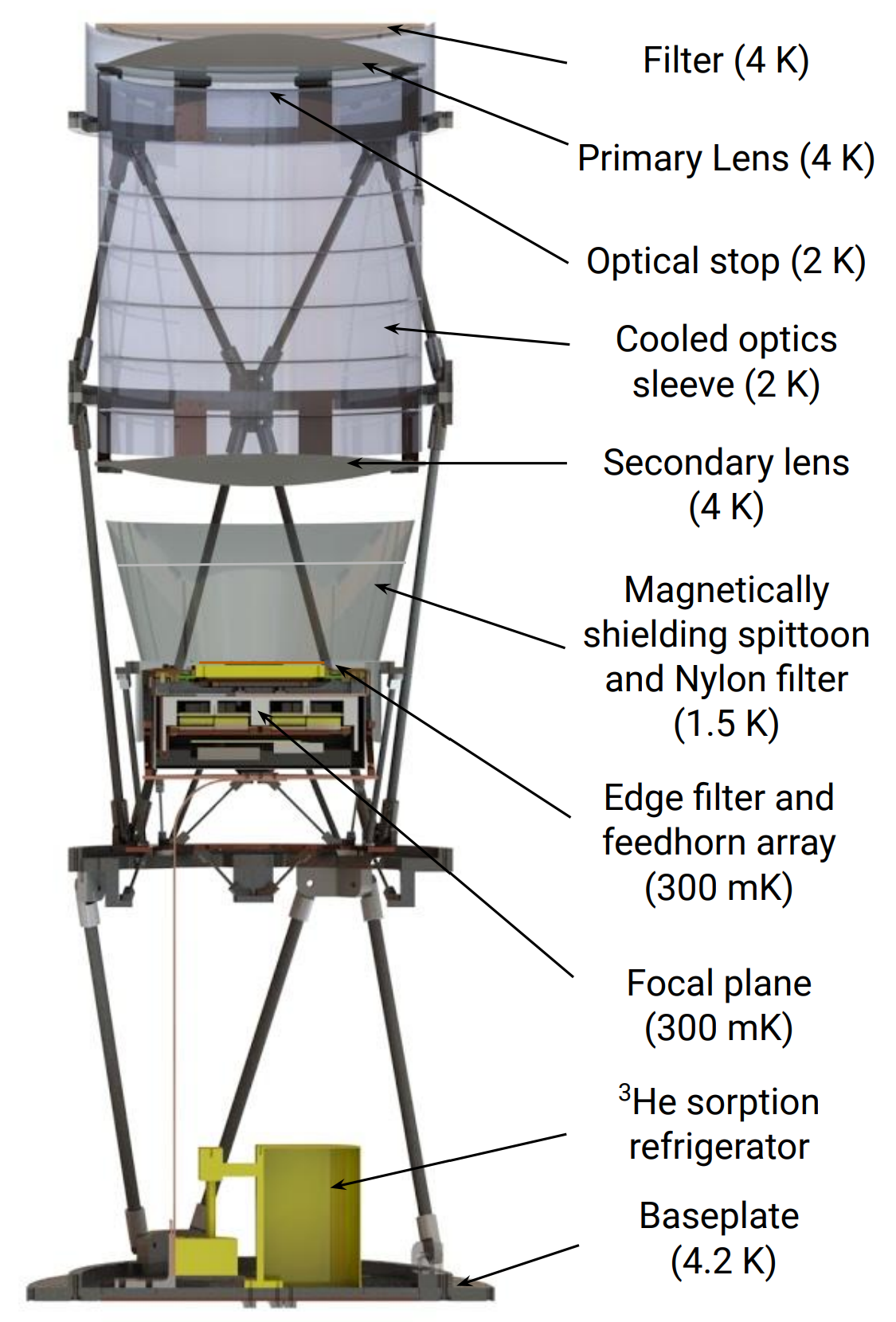}
\end{center}
\end{minipage}\hfill
\begin{minipage}[ht]{.4\linewidth}
\begin{center}
    \includegraphics[height=13cm]{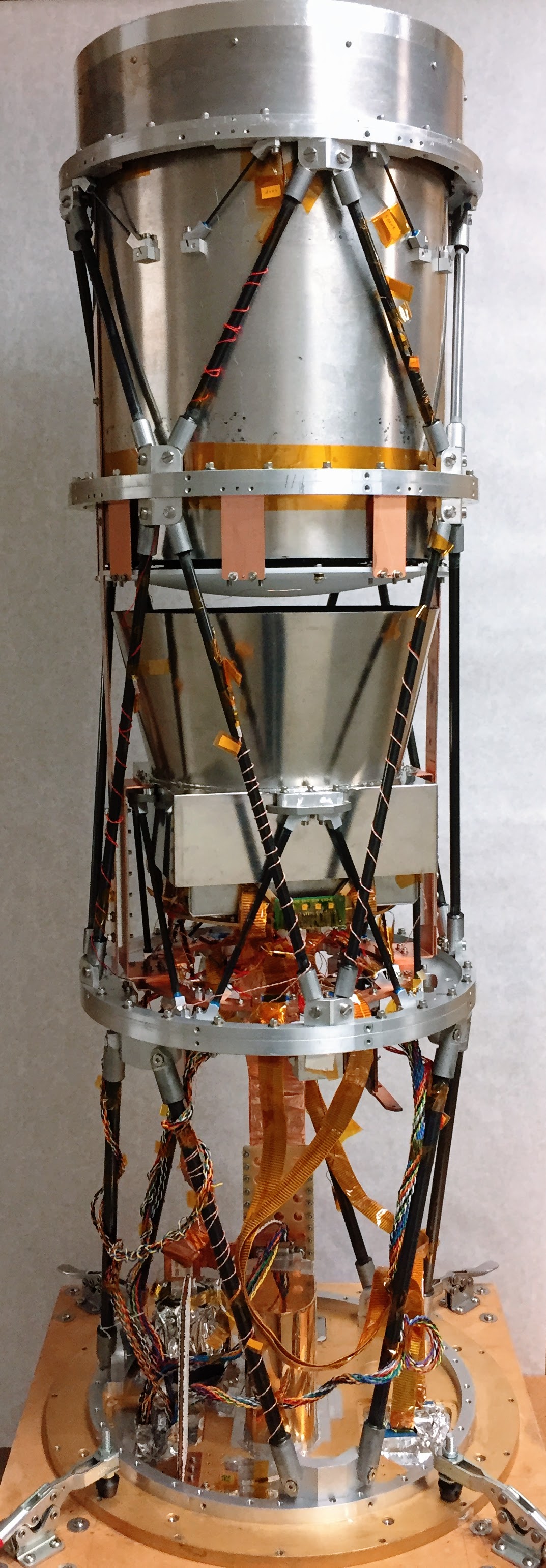}
\end{center}
\end{minipage}\hfill
\caption{\textit{(Left)}: A simplified cutaway CAD model of the \textsc{Spider}-2 280~GHz receivers, including the optics, baffling systems, filters, focal plane, and $^3$He sorption refrigerator. The 280~GHz optics are approximately 20~cm shorter than the 95- and 150~GHz optics, resulting in a greater distance between the focal plane and base plate in the 280~GHz receiver. In this CAD view the interior of the optics sleeve does not show the blackening material, and the SQUID series array modules and 2~K copper post are not visible. 
\textit{(Right)}: A fully integrated 280~GHz receiver on the bench between rounds of optical testing. 
Before installation into a cryostat the telescope is wrapped in copper-clad G-10 and taped at the seams to be light-tight. 
The 280~GHz receivers stand approximately 1.2~m ($\sim$ 47 in) tall when measured from the top of the base plate to the top of the 4K filter stack.
}
\label{fig:insert}
\end{figure}

The three new 280~GHz receivers have been rebuilt using some components from previously-flown 95- and 150~GHz receivers, well described in Ref.~\citenum{Runyan_2010,Rahlin_2014}, with design changes to accommodate the 280~GHz optics (see Figure~\ref{fig:insert}). 
Below we focus primarily on the new aspects of the 280~GHz receivers.
The 280~GHz (Rev-Y) FPUs received the nicknames `Y3', `Y4', and `Y5', based on the \spider-1 receiver from which their FPU packaging was derived.
We later refer to `Y4b', which is the second iteration of the Y4 FPU with a new detector wafer.
The motivation behind swapping out the Y4 detector wafer is discussed in \S\ref{sec:OpticalEff}.

\subsection{Receiver Architecture}
Each receiver is built up from a gold-plated aluminum cold plate bolted to the MT. 
The cold plate has feed-through ports for thermometry wiring, data cables, and for a gold-plated copper post that forms the thermal connection between the superfluid tank and the 2~K stages in the telescope.
Sub-kelvin cooling power for the detector arrays is provided by a single-stage helium-3 sorption refrigerator mounted to the base plate of each telescope.
A flexible, multi-layered copper shim heat strap provides strong thermal connection between the sorption cooler and the focal plane unit. Similar straps connect the refrigerator's condensation point and a 2~K heat-sink ring to the copper post. 
Two sets of solid copper heat straps provide thermal links from the 2~K ring to the magnetically-shielding focal plane enclosure and the cooled optics sleeve baffle located between the lenses.

The rigid structure for each receiver is provided by a lightweight frame composed of carbon fiber and aluminum.
The truss legs are carbon fiber rods epoxied into aluminum end cap fixtures, and are organized into three tiers separated by aluminum rings.
After the first few cryogenic cycles of the 280~GHz telescopes, we discovered that a number of the epoxy joints between the carbon fiber and aluminum end caps showed evidence of failure at the aluminum-epoxy interface.
These failures most often presented as hairline cracks in the epoxy (generally with no visible effect on the truss leg length), or as a larger crack with visible separation ($< 0.5$~mm) between the end cap and its original location.
Closer inspection of the \spider-1 telescopes showed similar epoxy joint issues on a few of the truss legs (which had undergone upwards of 20 cryogenic cycles), though
it is not apparent that the cracked epoxy joints had any impact on the instrument performance.
We believe poor surface preparation and ill-mixed epoxy batches to have been a large contribution to the bad epoxy joints on the new truss legs.
Out of caution we designed new end caps with interior grooves for the epoxy to flow into that should preserve a truss leg's length even with delamination at the epoxy-aluminum interface.
We replaced all of the truss legs on the 280~GHz telescopes, all compromised truss legs on the 95- and 150~GHz receivers, and we intend to replace the rest before flight.


\subsection{Focal Plane}
At the heart of each telescope is the detector focal plane unit (FPU), consisting of the detector arrays and sub-Kelvin readout electronics encased in magnetically shielding packaging.
The entire FPU is cooled to 300~mK by the $^3$He sorption fridge. 
The 95 and 150~GHz FPUs each support four wafers of dual-polarization, slot-antenna-coupled superconducting transition edge sensors (TESs) fabricated at the Micro-Devices Lab at the Jet Propulsion Laboratory~\cite{Ade_2015}.
Each 95~GHz (150~GHz) wafer supports 288 (512) TESs in a 100~mm diagonal square.
The new 280~GHz FPUs are populated with a single 150-mm square wafer of feedhorn-coupled superconducting TES arrays fabricated at NIST~\cite{Hubmayr_2016, Bergman_2018}.
While the higher observing frequency of the new 280~GHz arrays allows their detectors to be packed more densely, readout limitations in the existing \spider hardware cap the number of detectors at 512.
The 256 TES pairs thus take up a smaller focal plane area than for the lower-frequency arrays, allowing for optics optimization (see \S\ref{sec:optics}). 

\begin{figure}[tbp]
\centering
\begin{minipage}[!ht]{.3\linewidth}
\begin{center}
    \includegraphics[height=4.75cm]{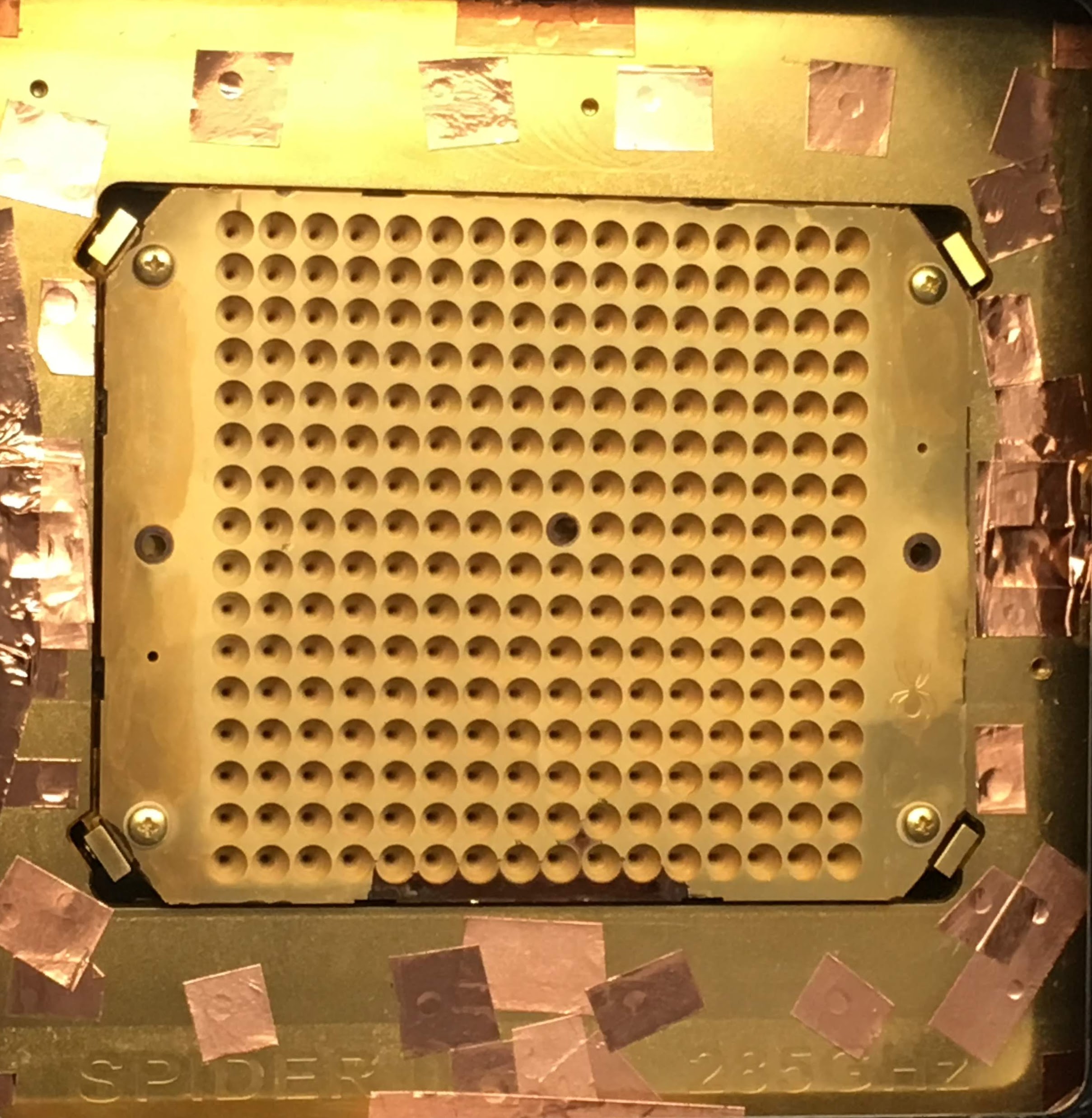}
\end{center}
\end{minipage}
\begin{minipage}[!ht]{.36\linewidth}
\begin{center}
    \includegraphics[height=4.75cm]{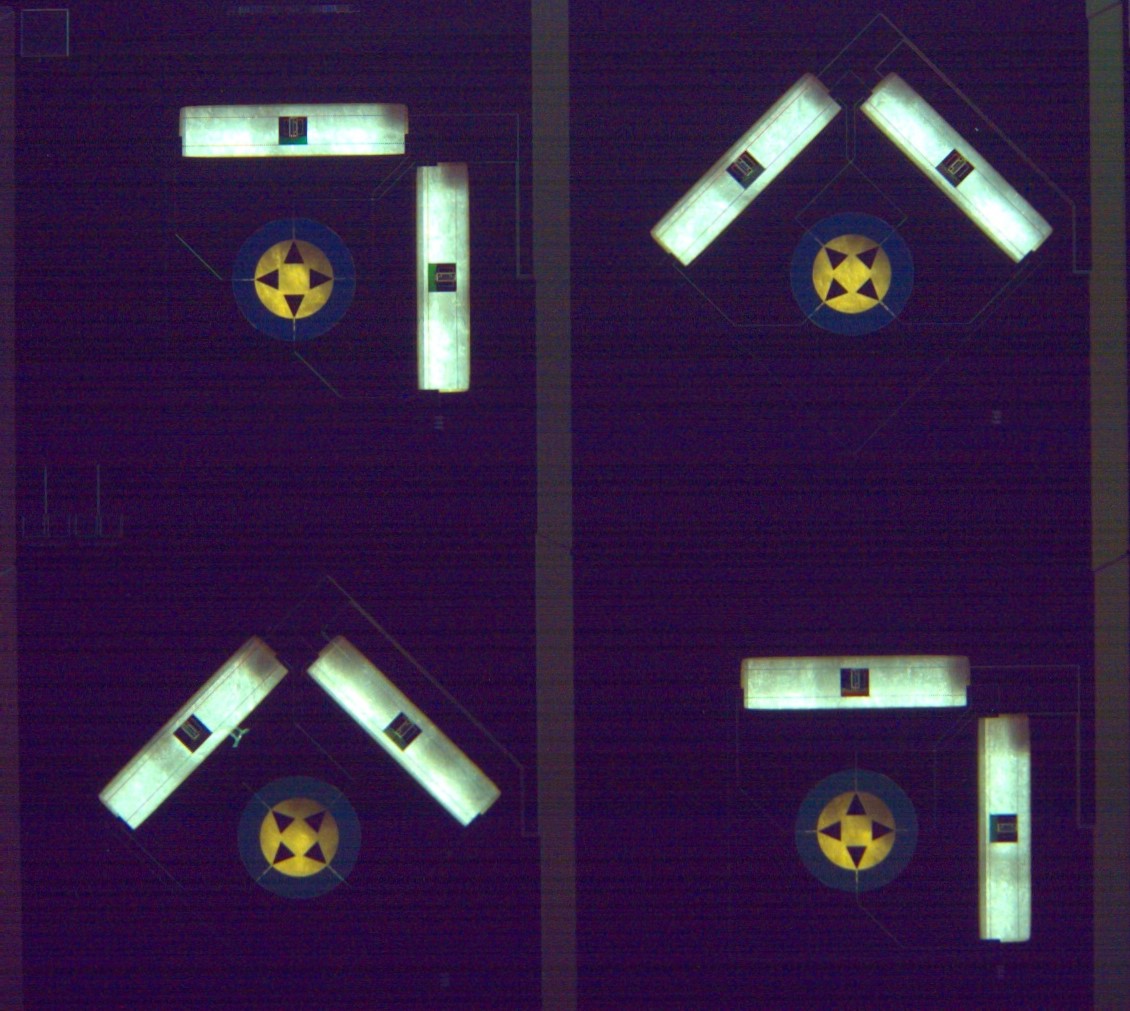}
\end{center}
\end{minipage}\hfill
\begin{minipage}[!ht]{.33\linewidth}
\begin{center}
    \includegraphics[height=4.75cm]{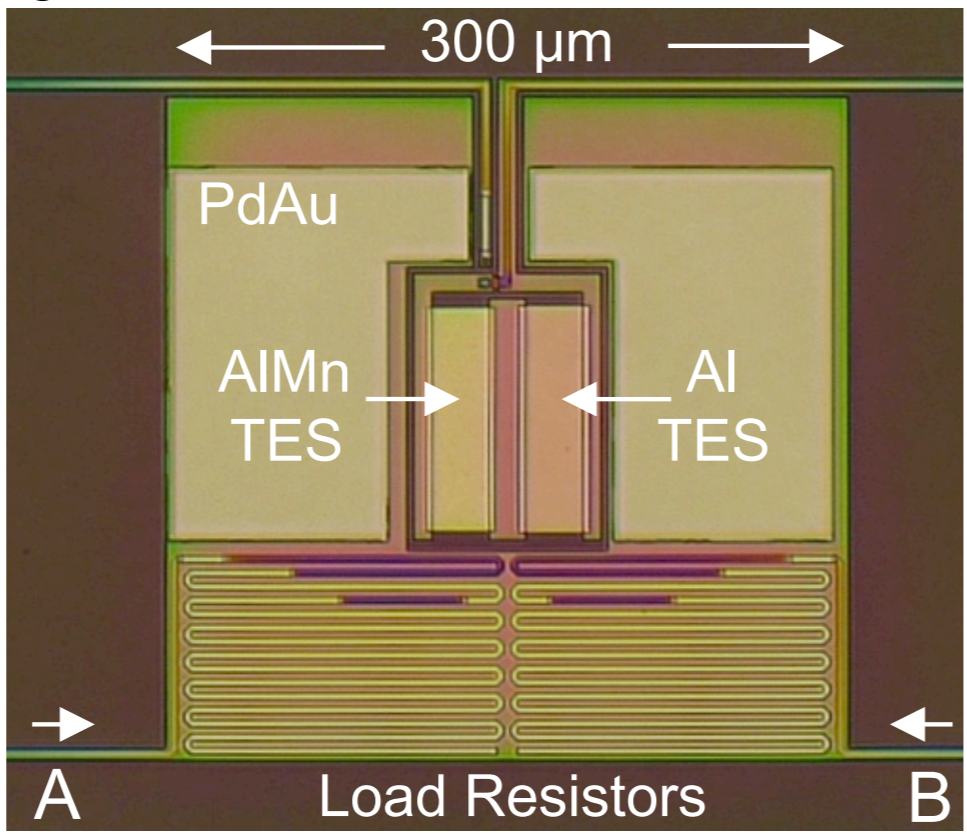}
\end{center}
\end{minipage}
\caption{
\textit{(Left)}: Feedhorn array on the `Y4b' 280~GHz focal plane. Each feedhorn directs power onto two TES islands through an orthomode transducer (OMT). The 315~GHz reflective metal-mesh low pass filter that mounts above the feedhorns is not shown. 
\textit{(Middle)}: Four TES pairs on the Y4b detector wafer showing the $0\degree$ and $45\degree$ alternation of neighboring pixels. Each pixel has two TES islands sensitive to orthogonal polarization modes. 
\textit{(Right)}: 280~GHz TES island.
Optical power is deposited on the island in the meandered gold load resistors.
Power dissipated by the voltage-biased TES fluctuates with its temperature-dependent resistance. 
The bolometer contains two sensors wired in series: a 420 mK AlMn sensor for science observations and a $\sim$ 1.6 K Al
sensor to extend the bolometer dynamic range for lab-based measurements.
Cooling power is provided from the thermal bath through weakly conductive legs suspending the island. 
Reproduced from Ref.~\citenum{Hubmayr_2016}.}
\label{fig:TESisland}
\end{figure}

\begin{figure}[ht]
\centering
\includegraphics[width=16cm]{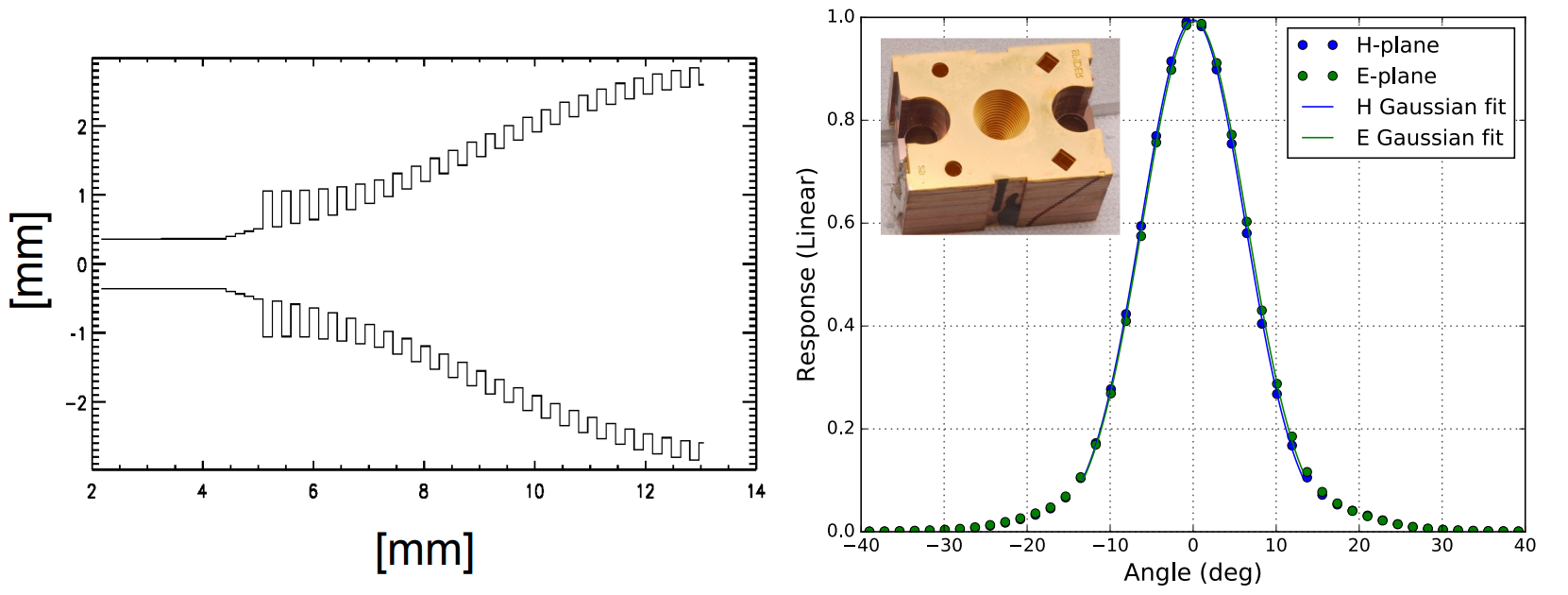}
\caption{\textit{(Left)}: Feedhorn corrugation profile.
\textit{(Right)}: Measurement and Gaussian fit to the E-plane and H-plane angular response of a single pixel feed, shown in the inset.
Points are the average measurement over the frequency band, and solid lines are the Gaussian fit. Reproduced from Ref.~\citenum{Hubmayr_2016}.}
\label{Spider2_FH}
\end{figure}

The monolithic, corrugated feedhorn arrays are made of Au-plated silicon; a photograph of a 280~GHz feedhorn array is shown in Figure~\ref{fig:TESisland}, and the feedhorn corrugation profile is shown in Figure~\ref{Spider2_FH}.
The feedhorns couple to pairs of planar orthomode transducers (OMTs), which direct orthogonal polarization modes to distinct TESs.
Nearest-neighbor bolometer pairs in the 280~GHz arrays are rotated by 45$\degree$ with respect to one another. 
This allows for each focal plane to simultaneously recover the Stokes Q and U parameters, before HWP rotation.
Figure~\ref{fig:TESisland} depicts the orientations of the spatial pixels and the architecture of the 280~GHz TES bolometer islands.


Every \spider bolometer has two TESs with different superconducting critical temperatures connected in series on a suspended bolometer island. For the 280~GHz detectors, an AlMn TES ($T_c\sim420$~mK) provides low noise for science data acquisitions, while an Al
TES ($T_c\sim1.6$~K) extends the bolometer dynamic range for the higher optical loading conditions of lab-based measurements.
Each bolometer island is thermally isolated from the rest of the detector wafer by weakly conductive silicon nitride legs.
Optical power directed from the OMTs is deposited on the bolometer islands in meandered gold-film resistors.
The power dissipated by the voltage-biased TES changes with its temperature-dependent resistance, and the associated change in current is detected by the cold readout electronics.
\spider's TES bolometers are tailored to take advantage of extremely low background loading in the space-like flight environment; device characterization is discussed in \S\ref{deviceparameters}.

The \spider TESs are read out using a time-division multiplexed Superconducting QUantum Interference Device (SQUID) system~\cite{SQUID}.
The TESs and first two stages of SQUIDs are encased inside the focal plane unit. The final readout stage, the SQUID series array (SSA), is located outside of the FPU.
In \spider's first flight, the SSAs for four of the six telescopes were mounted to the 4~K main tank. 
Since the MT empties during flight, the temperature of the MT surface was not sufficiently stable to mount the SSAs of the two highest telescopes.
For these, a newer SSA design was used that could be operated sunk to the more stable 2~K stage.
For \spider-2 all six telescopes use SSAs of this design mounted at the 2~K stage. 


In flight, \spider scans across the sky and moves through the Earth's magnetic field throughout its two-week trip around the South Pole.
Due to the sensitivity of SQUIDs and TESs to magnetic fields, this necessitates extensive magnetic shielding within the instrument.
Shielding is achieved through multiple layers of high-permeability and superconducting materials mounted within and around the focal plane unit. 
Most visible from the exterior are the superconducting niobium box surrounding the bottom of the FPU and enclosing the SQUID and SSA modules, and the high-permeability \textit{Amumetal A4K}\footnote{\textit{Amuneal Manufacturing Corp.}, Philadelphia, PA.} box that wraps around and extends upwards from the focal plane with a conical aperture. 
Laboratory measurements detailed in Ref.~\citenum{Bergman_2018} confirm that the 280~GHz detectors will be adequately shielded from transient magnetic fields during flight.

\subsection{Optical Design} 
\label{sec:optics}
All \spider receivers are cooled two-lens on-axis refracting telescopes.
Changes to the 280~GHz optics from the \spider-1 design stem from the physical size of the 280~GHz array, which is smaller than the 95- and 150~GHz focal planes while containing a similar number of sensors, and the relatively Gaussian beam profile and low sidelobe levels provided by the feedhorns (Figure~\ref{Spider2_FH}). 
This allows us to optimize the beam quality and edge taper over a smaller focal plane footprint.

The optical design was constrained to accommodate the size of the focal plane and beam shape, with a target far-field beam size of 17~arcminutes FWHM. 
Additional constraints were included to keep the curvature and thickness of the lenses at manageable values for fabrication and dielectric loss, respectively.
Optical simulations were conducted using the commercial software Zemax OpticStudio.
The resulting design for the 280~GHz receivers has a shorter effective focal length and smaller field of view than the 95 and 150~GHz optics.
The lenses are simple conics separated by 365 mm and have an effective focal length of 454.7 mm and a field of view of 16.4 degrees across the diagonal, compared to 550 mm lens separation, 583.5 mm effective focal length, and a 20 degree field of view for the 95- and 150~GHz optics\cite{Runyan_2010}.  
Changes to the telescope architecture to accommodate the new, shorter optical design included adjustments to the length of the carbon fiber truss legs, shortening the cooled optics sleeve, and modifying the copper heat straps to the 300~mK and 2~K stages to make up for the larger distance between the cold plate and the focal plane unit.


\subsection{Lenses and Anti-Reflection Coatings}
The \spider lenses are machined from high-density polyethylene (HDPE), and are cooled to 4~K to reduce in-band loading from dielectric loss in the plastic.
At millimeter wavelengths the index of refraction of the HDPE is $\sim 1.52$\cite{Lamb1996}.
We bond an anti-reflection (AR) coating layer to the surface of each lens to suppress reflections of order $\sim 4\%$ at each surface.
We use sheets of porous Teflon manufactured by Porex\footnote{\textit{Porex Filtration Group}, Atlanta, GA.} that is optimized for the 280~GHz band by choosing the Teflon density such that the index of refraction is close to the ideal $n_{AR} = \sqrt{n_{HDPE}}$. 
The sheet thickness is then chosen to be as close to $\lambda/4n_{AR}$ as possible, while being constrained by the availability of commercial materials. 
The same AR material is used to AR coat the vacuum windows and nylon filters used in the 280~GHz telescopes.
The selected 280~GHz AR material is $\sim0.18$ mm thick and has an index of refraction near 1.31, and has been chosen with AR performance on all three optical materials in mind. 

The HDPE lenses were fabricated by the Mechanical Engineering machine shop at the University of Illinois at Urbana-Champaign (UIUC).
Measurements of the lenses shape and curvature fall within the 5 mil machining tolerance.
The AR coats are bonded to the lens surfaces using a thin sheet of low-density polyethylene.
During the oven cycle, the lens and AR coating layers are placed under vacuum in order to prevent the formation of air bubbles underneath the AR coat.
One side of the lens rests on a surface custom fit to the lens curvature, and the other side is compressed with a sheet of silicone, with a layer of breather material between to allow adequate air flow when the vacuum is pulled. 
For the higher curvature lens surfaces, the AR material is pre-stretched before bonding to reduce wrinkling of the AR layer during the oven cycle. 

\subsection{Band Definition and Filtering}\label{sec:BandpassFilters}
Our desired observation bands are achieved with filtering on the detector arrays and optical elements in front of the detectors.
Unlike the 95 and 150~GHz arrays, the 280~GHz arrays do not have an on-chip band-defining filter.
The low-frequency edge of the band is defined by the cutoff frequency of the waveguide on the detector side of each feedhorn. 
The high-frequency edge is defined by a 10.5~cm$^{-1}$ ($\sim315$~GHz) low-pass metal-mesh filter~\cite{metal_mesh_review}.
In order to mitigate potential above-band light leaks, rather than mounting this filter at the magnetic shield aperture we have mounted it at 300~mK directly above the feedhorn array.

\spider utilizes a number of reflective, metal-mesh filters along the optical path to block unwanted infrared light while contributing minimal in-band loading to the detectors.
Three reflective ``shaders" mounted directly below the vacuum window bucket provide the first line of defense against infrared radiation. 
Four more shaders are mounted to $\sim$ 160~K and $\sim$ 40~K at VCS2 and VCS1, respectively. 
Two multi-layer hot-pressed metal-mesh low-pass filters are mounted immediately on the cold side of the VCS1 shaders.
Another 12~cm$^{-1}$ (360~GHz) cutoff filter is mounted at 4~K in the `snout' at the entrance of the telescope itself.
A single 3/32" anti-reflection coated nylon filter is mounted in the 1.6~K magnetic shield aperture to absorb additional infrared radiation.
Copper-clad G-10 wraps surround the entire receiver and are sealed with aluminum tape at the seams to be light- and RF-tight.

\subsection{Baffling and Optical Stop}
An optical baffling sleeve cooled to $\sim$2~K by the superfluid tank surrounds the space between the primary and secondary lenses to reduce optical loading onto the detectors.
The optics sleeve has rings extending into the interior and is blackened to absorb sidelobe power falling outside of the stop \cite{Runyan_2010, Rahlin_2014}.
The top ring in the cooled optics sleeve sits directly below the primary lens and forms the optical stop that limits beam spillover onto the warmer stages of the cryostat.

The optics sleeve is formed from soldered aluminum lined with Eccosorb HR-10 foam\footnote{\textit{Laird Technologies, Inc.}}, adhered with Stycast 2850 epoxy. 
We were able to reuse the optics sleeves from \spider-1 in the shorter focal length 280~GHz design by removing approximately 7 inches from each sleeve.
Zemax simulations demonstrate that keeping the baffling rings in the same location spacing as for the 95- and 150~GHz optics has no significant effect on the beam at the stop.

\subsection{Wave Plate}
The single-axis symmetry of the telescopes is beneficial for reducing polarized instrumental systematics that could show up in the data as false sky signal.
To mitigate beam systematics and to increase polarization coverage, sky polarization modulation is provided by a monochromatic half wave plate (HWP)~\cite{Bryan_2010}. 
The HWPs are rotated twice per sidereal day in multiples of 22.5 degrees according to a pre-determined schedule. We optimize the schedules for individual HWPs according to their orientation in the flight cryostat.
The 280~GHz HWPs are 1.66~mm thick birefringent sapphires from Rubicon Technology\footnote{\textit{Rubicon Technology}, Bensenville, IL.} with a $\sim$0.125~mm Cirlex AR coat\cite{Bryan_2016}.  
The HWP rotation mechanisms are mounted directly onto the MT in front of the primary lenses and 4~K filter. 

\subsection{Vacuum Window}
\begin{figure} [ht]
\centering
\begin{minipage}[ht]{.5\linewidth}
\begin{center}
    \includegraphics[height=5cm]{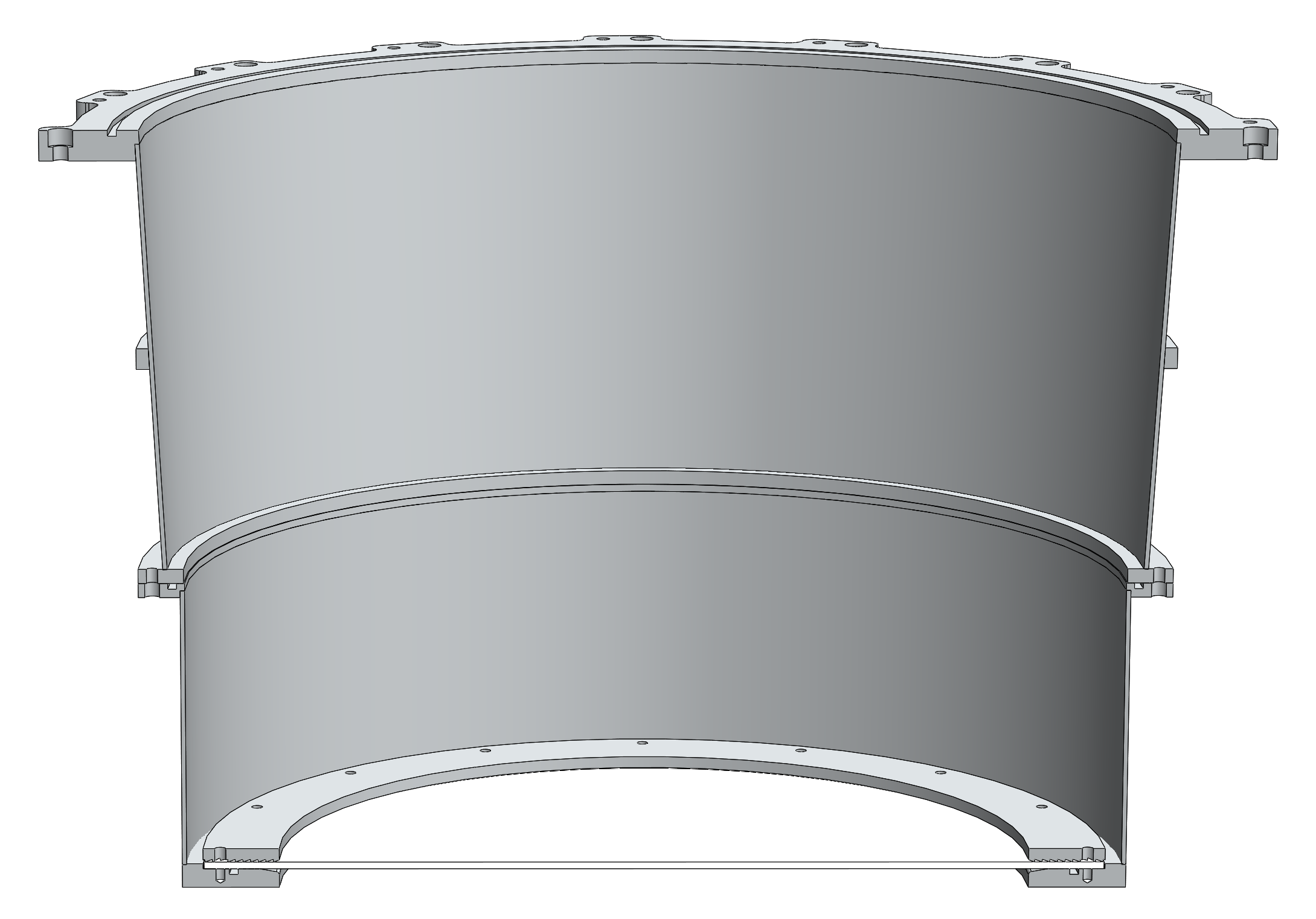}
\end{center}
\end{minipage}\hfill
\begin{minipage}[ht]{.5\linewidth}
\begin{center}
    \includegraphics[height=5cm]{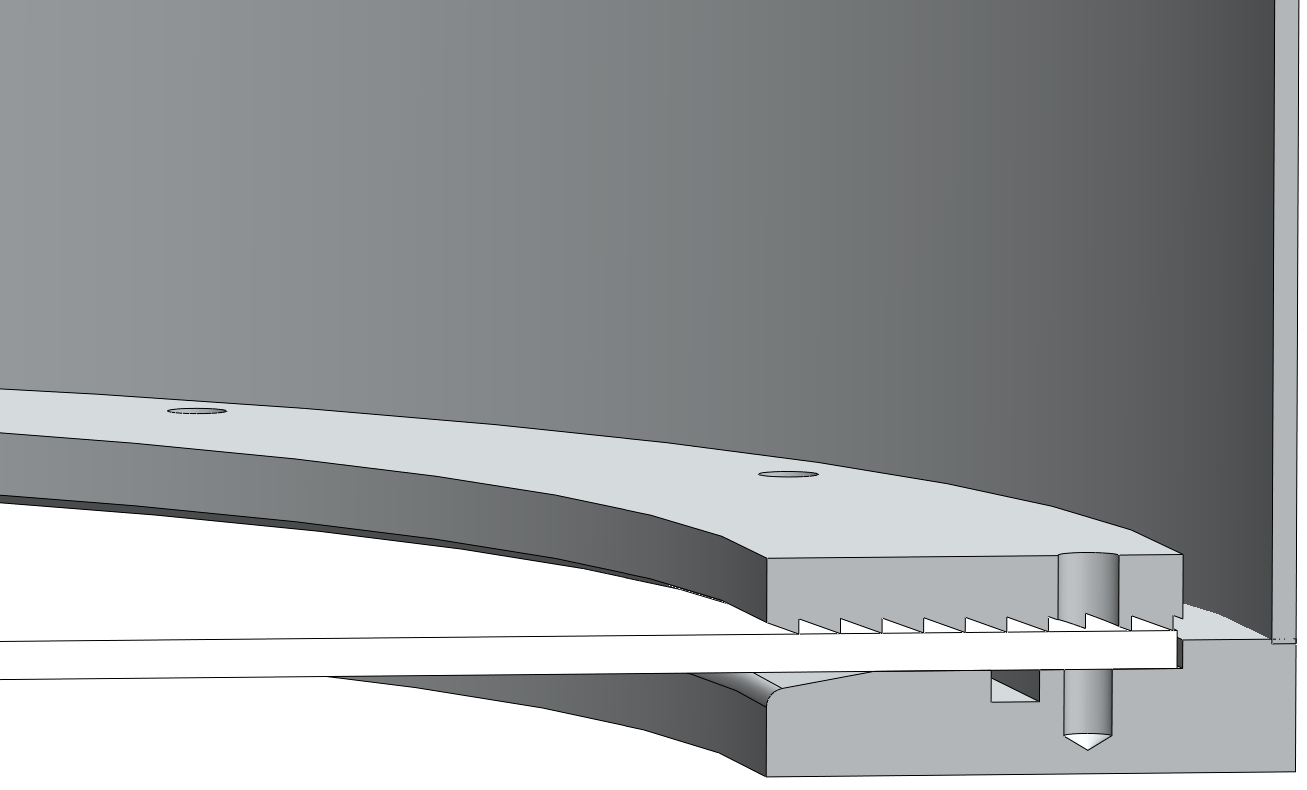}
\end{center}
\end{minipage}
\caption[example] 
{ \label{fig:WindowClamp} 
{\textit{Left}}: Cross-sectional view of the window bucket that holds the UHMWPE window, here depicting a flat non-vacuum cycled window.  {\textit{Right}}: A closer view of the \spider vacuum window clamping mechanism. The top ring has milled concentric teeth to grip the window. The bottom ring has an o-ring groove to form the vacuum seal and a beveled edge to reduce strain on the plastic window as it pulls inward under vacuum.} 
\end{figure}

The 340~mm aperture vacuum windows are AR-coated, ultra-high molecular weight polyethylene (UHMWPE) plastic sheets that are transparent to millimeter wavelengths. 
The windows are held in place between two aluminum clamping rings at the bottom of the window buckets.
The upper ring screws into threaded holes in the lower ring.
The low coefficient of friction of the UHMWPE makes the windows difficult to hold in place, so the upper ring is milled with concentric teeth that grip into the plastic.
The lower ring has a groove for an o-ring to hold vacuum, and a smooth chamfered edge to reduce strain on the UHMWPE plastic when under vacuum (see Figure~\ref{fig:WindowClamp}).
\spider-1 utilized 1/8"-thick windows for all the telescope apertures.
We have chosen to use 1/16" windows in front of the 280~GHz telescopes to reduce optical loading from the UHMWPE at high frequencies.
The thin UHMWPE windows plastically deform while holding a vacuum. 
Before installation into the flight cryostat, each window is installed in a test vacuum chamber and is slowly stretched to test for leaks and to ensure the AR coating does not peel.
Window creep and deformation is closely monitored during the testing to ensure the window will not extend into the infrared blocking filters below the windows in the flight cryostat.

\section{Instrument Performance}\label{section:InstrumentPerformance}
\subsection{Detector Performance}
The 280~GHz focal planes and fully-integrated receivers have undergone extensive testing in the flight cryostat and in smaller test cryostats.
The performance and properties of the Y3, Y4, and Y5 arrays have been described in previous publications\cite{Hubmayr_2016, Bergman_2018}.
Here we update measurements for Y3 and Y5 and report on the properties of Y4b, the iteration of the FPU that will be deployed alongside Y3 and Y5.

\subsubsection{Bolometer parameters}\label{deviceparameters}
Each of the 512 TESs on a detector array has unique electrical, thermal, and optical properties that must be characterized before deployment.
Arrays are initially characterized ``dark'', i.e. with the optical chain blocked above the feedhorns with a gold-plated silicon wafer at 300~mK. 
We acquire I-V curves or ``load curves" at a range of bath temperatures; the bias current is swept and the output TES current is recorded.
When the TESs are on their superconducting transitions in the absence of optical power input, the Joule power dissipated in the TES by the bias circuit is balanced by the power dissipated through the thermal link to the bath and is equivalent to the saturation power ($P_{sat}$) --- the power required to heat the TES up to its critical temperature.
When the absorbed power on the bolometer island surpasses $P_{sat}$, the TES is driven normal and can no longer detect light.
As the bath temperature ($T_{bath}$) approaches $T_c$, $P_{sat}$ decreases.
%
We measure $P_{sat}$ for each detector at bath temperatures between 330~mK and 480~mK.
We obtain the detector parameters  $T_c$, thermal conductivity $G_c$ at the critical temperature, and the thermal sensitivity $\beta$ by fitting $P_{sat}$ and $T_{bath}$ data to the functional form 
\begin{equation}
\label{eqn:Psat}
    P_{sat}(T_{bath}) = K(T_c^{\beta+1}-T_{bath}^{\beta+1}),
\end{equation}
where $K = G_c/(\beta+1)T_c^{\beta}$, and we extrapolate down to the flight operational temperature, 300~mK.

Parameter measurements for the three deployment arrays are listed in Table~\ref{tab:FPUParameters}.
We find high uniformity in detector parameters across the arrays, with small variations due to fabrication gradients. 
The target saturation power ($P_{sat}$(300~mK)) of the 280~GHz TESs was 3~pW; the measured values for Y3 and Y5 are close to this specification.
The Y4b array has a higher median saturation power and critical temperature, which does not significantly diminish the performance of the detectors.

\subsubsection{Yield}
The end-to-end channel yield is estimated from analysis of load curves taken at base temperatures.
We achieved high yield on all three 280~GHz arrays.
The final yield is approximately 86\%, 86\%, and 95\% for Y3, Y4b, and Y5, respectively (Table~\ref{tab:FPUParameters}).
The dominant defects are from broken wiring and SQUID amplifier failures. 

\begin{table}[ht]
\caption{End-to-end yield and device parameter estimates for the three 280~GHz focal planes. Parameters were measured at various bath temperatures from 330 -- 480~mK, and extrapolated down to the flight operating temperature at 300~mK (\S\ref{deviceparameters}) where appropriate. Listed for each array is the median detector saturation power, critical temperature, thermal conductivity at the critical temperature, normal resistance, thermal conductance power law (Eqn.~\ref{eqn:Psat}), and detector yield. The variation in measured parameter values is reported at 1$\sigma$. Values for Y3 and Y5 are reported from Ref.~\citenum{Bergman_thesis}. The naming convention of the 280~GHz focal planes stems from the \spider-1 FPU from which their packaging came.}
\label{tab:FPUParameters}
\begin{center}       
\begin{tabular}{|c|c|c|c|c|c|c|} 
\hline
\rule[-1ex]{0pt}{3.5ex}  FPU & $P_{sat}$(300 mK)& $T_c$& $G_c$ & $R_n$  &$\beta$ & Yield\\
& [pW]&[mK]&[pW/K]&[mOhm]&&[\%]\\
\hline
\hline
\rule[-1ex]{0pt}{3.5ex}  Y3 & 2.76 $\pm$ 4.8\% & 440   $\pm$ 0.8\%  & 27.74   $\pm$ 4.6\% & 10.91 $\pm$ 2.9\% & 1.88 $\pm$ 1.5\% & 86\%\\
\hline
\rule[-1ex]{0pt}{3.5ex}  Y4b &  4.18    $\pm$ 4.6\% & 460   $\pm$ 0.9\%  & 36.12   $\pm$ 4.3\% & 11.54 $\pm$ 2.7\% & 1.76 $\pm$ 7.0\% & 86\%\\
\hline
\rule[-1ex]{0pt}{3.5ex}  Y5 &  3.54    $\pm$ 4.8\% & 450   $\pm$ 0.4\%  & 33.00   $\pm$ 4.7\% & 10.68 $\pm$ 3.6\% & 2.03 $\pm$ 0.9\% & 95\% \\
\hline
\end{tabular}
\end{center}
\end{table}


\subsection{Optical Performance}
The 280~GHz telescopes have been tested for optical efficiency, spectral response, near-field beam mapping, and internal loading on the detectors. 
Optical testing of the completed telescopes is largely performed on the high-$T_c$ lab transition, that does not saturate easily in high-loading environments.
In order to provide sufficient Joule power to the lab TESs, we needed to modify the bias circuitry for laboratory testing.

\subsubsection{Bands}
Observations at 280~GHz will complement \spider's 95 and 150~GHz data sets by providing necessary data to characterize Galactic dust emission.
The 280~GHz \spider observation band falls between the \planck-HFI 217 and 353~GHz bands, and avoids the rotational transition spectral lines from Galactic CO near those frequencies\cite{collaboration2018planckoverview}.

\begin{figure}[ht]
\centering
\begin{center}
    \includegraphics[height=7.5cm]{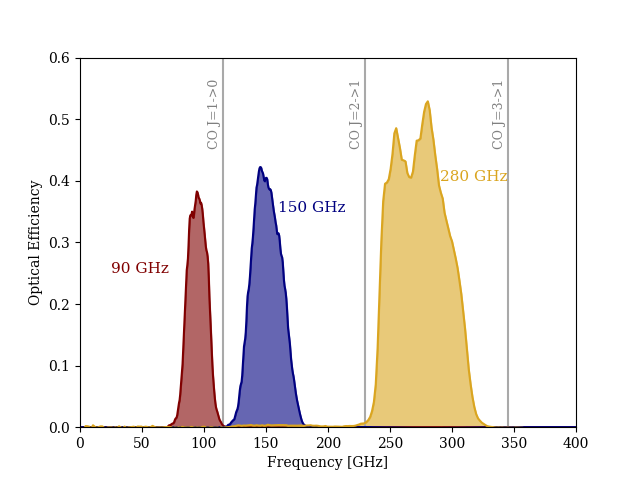}
\end{center}
\caption{End-to-end frequency response of the three \spider bands shown as the mean across detectors. 95- and 150~GHz bands are as measured in \spider-1 flight cryostat before flight, and 280~GHz as measured in the \spider-2 flight cryostat. Spectral response is averaged across detectors and normalized to reproduce the measured optical efficiencies.}
\label{Bands}
\end{figure}

We have measured the spectral response of the Y3 and Y5 receivers using a Martin-Puplett interferometer mounted at the aperture of the \spider-2 flight cryostat. 
Figure~\ref{Bands} shows mean bandpasses for detectors in the Y3 and Y5 arrays, 
alongside similar measurements for the 95 and 150~GHz receivers taken in the \spider-1 flight cryostat during the final characterization tests before flight.
In the figure, the spectral responses have been normalized to reproduce the measured optical efficiencies (discussed in \S\ref{sec:OpticalEff}).
The band centers and bandwidths for all three \spider frequencies are detailed in Table~\ref{tab:BeamSummary}. The median bandwidth for the measured 280~GHz channels is 71.2 GHz, centered at 270.2 GHz.

\subsubsection{Beams}
In order to get an initial sense of the beam quality of the 280~GHz detectors, we have mapped the near-field response of the Y4b 280~GHz telescope.
A near-field beam map (NFBM) was taken at the aperture of the UIUC test cryostat using a chopped source and two-dimensional translation stage. 
The source consists of an amplified noise source in the Y-band with a nearly flat spectrum across the receiver band. 
The source signal passes through a waveguide attenuator to avoid saturating the aluminum TESs. 
The source is coupled to free space by means of a feedhorn and is positioned 
a few inches above the vacuum window.
In this configuration the effect of the source beam is negligible, so the measured pattern is a good approximation to the telescope's near-field beam. 

\begin{figure}[ht]
\centering
\begin{minipage}[ht]{.5\linewidth}
\begin{center}
    \includegraphics[width=\textwidth]{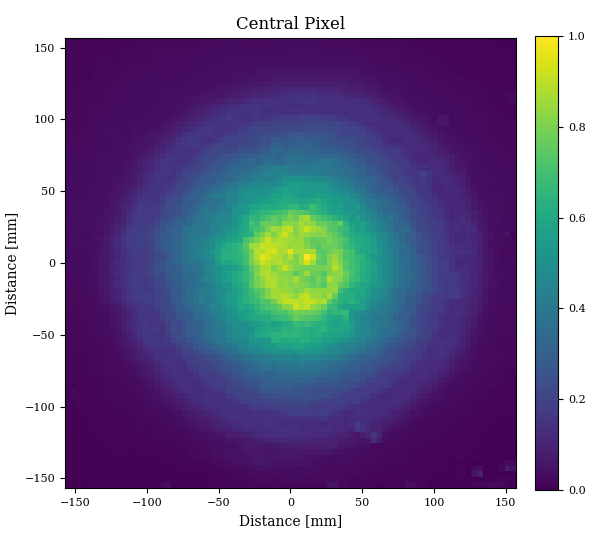}
\end{center}
\end{minipage}\hfill
\begin{minipage}[ht]{.5\linewidth}
\begin{center}
    \includegraphics[width=.95\textwidth]{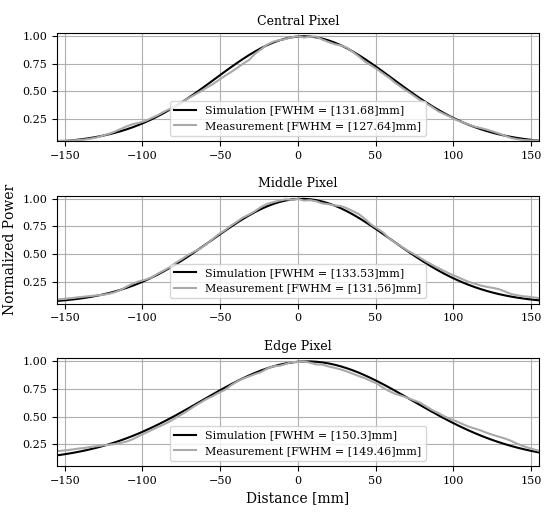}
\end{center}
\end{minipage}
\caption{\textit{(Left)}: Near-field beam map of a detector in the center of the Y4b \spider 280~GHz detector array.
Data was collected on a fully integrated 280~GHz receiver by scanning a source across the aperture of the test cryostat.
{\textit{(Right)}}: 
Comparison of measured beam cross section to Zemax physical optics simulations along one scanning direction.
Data for three detectors are shown representing central, middle range, and edge spatial pixel locations. 
The FWHM of the beams calculated in the plane of the NFBM scanning stage are shown for both simulation and measurement.
}
\label{NFBM_cent}
\end{figure}

The source radiation is modulated at 8~Hz and synchronized with the acquisition electronics so that the detector response can be reconstructed offline by means of a synchronous demodulation algorithm implemented in software.
Figure~\ref{NFBM_cent} (left) shows the normalized beam pattern of a detector located near the center of the Y4b array. 
The right panel compares cross sections of the observed beam profile to Zemax physical optics simulations of the optics chain with a Gaussian input beam at the horn location.

\begin{table}[ht]
\caption{
Summary of instrument parameters for \spider's second flight.
Listed values are the nominal telescope bands, the telescope center frequencies and bandpass, beam width at FWHM, total number of spatial pixels, total number of light-sensitive detectors, detector yield, and per-detector sensitivity.
Parameters reported for the 95- and 150~GHz bands are for the three receivers that will be redeployed in \spider-2.
The values reported are as measured in-flight or in the final pre-flight characterization of \spider-1.
Parameters listed for the 280~GHz band are the expected values for flight as measured in the lab, or as expected from models (noted with an *).
Band center and band width are averages of per-detector measurements, and beam full-width at half-maximum is extracted from a combined fit to all detectors in each band.
Single detector sensitivity is the mean noise-equivalent temperature per detector for the 95- and 150~GHz bands, and is a model based estimate for the 280~GHz band.
}
\label{tab:BeamSummary}
\begin{center}       
\begin{tabular}{|c|c|c|c|c|c|c|c|} 
\hline
\rule[-1ex]{0pt}{3.5ex}Band & Band Center & Bandpass & FHWM & Number of  & Number of & Yield  &NET$_\text{det}$  \\
& [GHz]& [\%]& [arcmin]& Spatial Pixels &  Detectors & [\%] & [$\mu \text{K}\sqrt{s}$] \\
\hline
\hline
\rule[-1ex]{0pt}{3.5ex}95& 94.9 & 26.5 & 41.3 & (2x) 144 & 576 & 83\% & 179.7 \\
\hline
\rule[-1ex]{0pt}{3.5ex}150& 151.4 & 25.2 & 28.8 & (1x) 256 & 512& 85\% & 157.9 \\
\hline
\rule[-1ex]{0pt}{3.5ex}280& 270.2 & 26.4 & 17* & (3x) 256  & 1536 & 89\% & 308*\\
\hline
\end{tabular}
\end{center}
\end{table}


\subsubsection{Optical efficiency}\label{sec:OpticalEff}
We measure the end-to-end optical efficiency of the receivers, including loss from the window, filters, and optics.
The optical responsivity is measured by comparing the Joule power of the detectors when presented with aperture-filling black body sources at room temperature (295~K) and at 77~K.
For our black body loads we use 
a foam dish filled with microwave-absorbing Eccosorb HR-25 foam backed by an aluminum metal plate.
We take I-V curves on the lab transition with the dish at room temperature, and again after filling the dish with liquid nitrogen.
We flush the air space between the dish and vacuum window with gaseous nitrogen to minimize water condensation onto the window.
The bands in Figure~\ref{Bands} are normalized to match this response.
A convenient figure of merit for optical efficiency is calculated as 
\begin{equation}
    \text{Optical Efficiency} = \frac{P_{sat}(295 K) - P_{sat}(77 K)}{295 K - 77 K}\frac{1}{dP/dT_{\text{ideal}}},
\end{equation}
where $dP/dT_{\text{ideal}}$ is the ideal single-moded detector response for a flat band of the specified bandwidth. 
The target optical efficiency for the 280~GHz detectors is near 40\%.

\begin{figure}[ht]
\centering
\begin{minipage}[ht]{.5\linewidth}
\begin{center}
    \includegraphics[width=\linewidth]{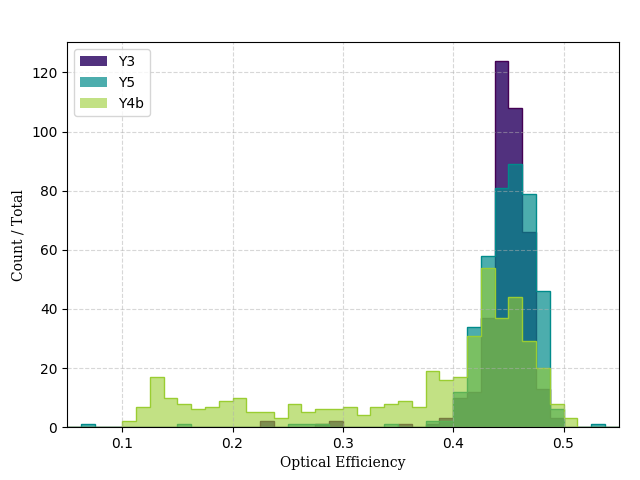}
\end{center}
\end{minipage}\hfill
\begin{minipage}[ht]{.5\linewidth}
\begin{center}
    \includegraphics[width=\linewidth]{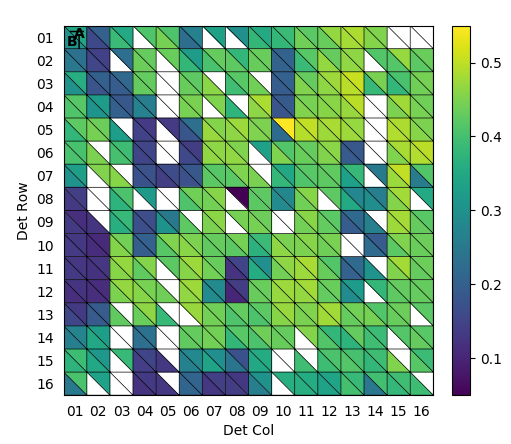}
\end{center}
\end{minipage}
\caption{
\textit{(Left)}: Histogram of the fractional end-to-end optical efficiency for the three 280~GHz receivers.  
Detector optical efficiencies on the Y3 and Y5 detector arrays is tightly distributed around 45\%, and the Y4(Y4b) array has a tail of low-efficiency detectors.
The low optical efficiency detectors are useful for characterizing the instrument under high optical loads.
\textit{(Right)}: End-to-end optical efficiency of the Y4b focal plane plotted in physical coordinates.
Physical location of low-efficiency detectors remained constant after the detector wafer swap. 
}
\label{fig:opticalefficiency}
\end{figure}

Figure~\ref{fig:opticalefficiency} shows a histogram of this end-to-end optical efficiency measure for all detectors on the three 280~GHz receivers.
Y3 and Y5 show tight distributions near 45\%, comfortably exceeding our design target.
Characterization of Y4, however, found that nearly a third of its detectors lie in a long tail extending to efficiencies as low as $\sim$10\%.
These low-responsivity detectors appeared in small clusters across the array, rather than following typical fabrication gradients.
These clusters were not visible in the detector parameters from dark characterization, which are relatively uniform across the array.
In order to rule out fabrication errors in the TESs or OMTs, the detector wafer alone was replaced, renamed `Y4b', and retested both dark and optically.
The tail of low-efficiency detectors remained (Figure~\ref{fig:opticalefficiency}), and in the same locations.
This suggests that the issue is a property of the feedhorn array.
The cost in mapping speed is minor enough that we plan to field Y4b without further modifications.

\subsubsection{Dark channels and radiation leakage}
One of the central feedhorn locations was used for alignment of the silicon layers during fabrication, leaving the 
two TESs underneath uncoupled to incident radiation.
One TES is intentionally disconnected from its SQUID readout, while the other is a fully-functioning detector, referred to as the ``dark TES''.
Both are useful for checking systematics~\cite{Hubmayr_2016}.

We use the dark TES and our optical efficiency measurements to make an estimate on the leakage radiation between detectors.
During our optical efficiency measurements, we find that the dark TES responds at a level 1.5\% of that  of the nearby optically-active TESs.
This exceeds the known $\sim$0.4\% crosstalk from the readout electronics seen in the disconnected channel.
This suggests a general magnitude for light leakage within the array, but since we do not stimulate one detector at a time (e.g. with a far-field beam map) it is difficult to pinpoint a specific leakage mechanism.

\subsubsection{Internal loading}
A final key parameter for the performance of these receivers is their internal loading: the radiation absorbed by the detectors due to emission from the receiver itself.
This may be due to in-band emission from receiver components, or to out-of-band radiation that leaks through the filter stack and reaches the bolometer.
Low internal loading is particularly critical for a balloon-borne instrument, due to the low atmospheric loading.
An elevated internal loading limits the high mapping speeds achievable with low-$G$ TESs in this environment, and if high enough can even saturate the detectors entirely.

\spider achieved very low internal loading during its first flight: $P_{\text{internal}}\lesssim$ 0.25~pW ($\lesssim$ 0.35~pW) absorbed at 95~GHz (150~GHz).
Insofar as the 280~GHz receivers inherit their design from these 95 and 150~GHz receivers, we would \emph{a priori} expect low loading.
A simple model of emission from the 280~GHz optical elements suggests $P_{\text{internal}}\lesssim$ 1~pW.
It can be difficult to directly constrain the internal loading of a balloon experiment in a lab environment, however, due to the difficulty of mimicking the radiation environment at float altitudes.
Early \spider-1 testing used an aperture-filling liquid helium cold load to estimate internal loading, and even then overestimated the in-flight internal loading significantly.

We have attempted to constrain the 280~GHz internal loading in the lab by taking advantage of the low-efficiency detectors on the Y4b focal plane, which are able to operate on the science transition under higher optical loads.
We reduce the optical loading environment by placing a curved mirror in front of the aperture, redirecting the detectors' beams back into the cold interior of the telescope and reducing spillover onto the warmer cryogenic stages.
Similar measurements of the 95 and 150~GHz receivers prior to the first flight yielded apparent loadings 3--5 times higher than those seen during flight. 
Our initial attempts to constrain internal loading on the 280 GHz receivers using this method suggested an apparent loading equivalent to a $\sim$30~K black body, however --- difficult to reconcile with plausible beam spills and emissivities.
This measurement could be increased by several effects: multiple reflections within the receiver, beam spillover, water condensation on the window coating, etc.
At the time of this writing, the \spider-1 cold load is being refit to attempt a similar measurement of the 280~GHz receivers.
This will provide a stronger constraint on internal loading, as well as a noise measurement on the 280~GHz detectors.

\section{Conclusions}
Three new 280~GHz receivers have been built and integrated into a new flight cryostat and gondola for the second flight of the \spider instrument from McMurdo Station, Antarctica.
The telescopes have undergone substantial testing in the flight payload, and are in final stages of optical characterization.
The three 280~GHz telescopes will be flown alongside two 95~GHz and one 150~GHz telescopes to reobserve a large portion of the southern Galactic sky with a total of over 2600 detectors in an effort to detect the primordial gravitational wave B-mode signal at degree angular scales.
The addition of the 280~GHz frequency band will improve \spider's ability to characterize and remove Galactic dust emission foregrounds from the CMB signal, as well as producing maps of polarized dust emission over a large sky area of lasting value to the field.

\acknowledgments 
\spider is supported in the U.S. by the National Aeronautics and Space Administration under grants\\ NNX07AL64G, NNX12AE95G, and NNX17AC55G issued through the Science Mission Directorate and by the National Science Foundation through PLR-1043515.
Logistical support for the Antarctic deployment and operations was provided by the NSF through the U.S. Antarctic Program. 
Support in Canada is provided by the Natural Sciences and Engineering Research Council and the Canadian Space Agency.
Support in Norway is provided by the Research Council of Norway.
Support in Sweden is provided by the Swedish Research Council through the Oskar Klein Centre (Contract No. 638-2013-8993).
The Dunlap Institute is funded through an endowment established by the David Dunlap family and the University of Toronto. 
K.F. is Jeff \& Gail Kodosky Endowed Chair in Physics at the University of Texas at Austin and is grateful for support. 
K.F. acknowledges support by the Swedish Research Council (Contract No. 638-2013-8993) and from the U.S. Department of Energy, grant DE-SC007859.
We also wish to acknowledge the generous support of the Lucile Packard Foundation, which has been crucial to the success of this project. 
The collaboration is grateful to the British Antarctic Survey, particularly Sam Burrell, for invaluable assistance with data and payload recovery after the 2015 flight. 
We thank Brendan Crill and Tom Montroy for significant contributions to \spider's development.

\bibliography{report} 
\bibliographystyle{spiebib} 

\end{document}